\input harvmac
\input amssym.tex

\def\newdate{CERN, 22/3/2004}


\def\a{\alpha}
\def\b{\beta}  
\def\g{\gamma}
\def\l{\lambda}
\def\d{\delta}
\def\e{\epsilon}
\def\t{\theta}

\def\D{\Delta}
\def\p{\partial}


\Title{
\vbox{
\hbox{YITP-SB-04-08}
}}   
{\vbox{
\centerline{Gauging Cosets}
}}  
 
\medskip\centerline
{P.~A.~Grassi$^{~a,b,}$\foot{pgrassi@insti.physics.sunysb.edu},
and  P.~van~Nieuwenhuizen$^{~a,}$\foot{vannieu@insti.physics.sunysb.edu}
} 
\medskip   
\centerline{$^{(a)}$ 
{\it C.N. Yang Institute for Theoretical Physics,} }  
\centerline{\it State University of New York at Stony Brook,   
NY 11794-3840, USA}  
\medskip
\centerline{$^{(b)}$ {\it Dipartimento di Scienze,
Universit\`a del Piemonte Orientale,}}
\centerline{\it
C.so Borsalino 54, Alessandria,  15100, ITALY}
\vskip .3cm  

\medskip  
\vskip  .5cm  
\noindent

\bigskip
\bigskip

\centerline{\bf Abstract}

\bigskip

We show how to gauge the set of raising and lowering generators 
of an arbitrary Lie algebra. We consider $SU(N)$ as an example. 
The nilpotency of the BRST charge requires constraints 
on the ghosts associated to the raising and lowering generators. 
To remove these constraints we add further ghosts and 
we need a second BRST charge to obtain nontrivial cohomology. 
The second BRST operator yields a group theoretical explanation 
of the grading encountered in the covariant quantization of superstrings.  

\Date{\newdate}


\listtoc
\writetoc

\newsec{Introduction}

\lref\ganor{
K.~Gawedzki and A.~Kupiainen,
Phys.\ Lett.\ B {\bf 215}, 119 (1988);
E.~Witten,
Nucl.\ Phys.\ B {\bf 371}, 191 (1992); 
O.~Aharony, O.~Ganor, J.~Sonnenschein, S.~Yankielowicz and N.~Sochen,
Nucl.\ Phys.\ B {\bf 399}, 527 (1993)
[hep-th/9204095]; 
O.~Aharony, O.~Ganor, J.~Sonnenschein and S.~Yankielowicz,
Nucl.\ Phys.\ B {\bf 399}, 560 (1993)
[hep-th/9208040].
} 

\lref\stora{
S.~Ouvry, R.~Stora and P.~van Baal,
Phys.\ Lett.\ B {\bf 220}, 159 (1989).
}

Nonlinear sigma models based on group or coset manifolds 
allow one to construct interacting string models with  
nontrivial backgrounds.  Three 
classes of models have been obtained in this way: 
{\it i)} standard WZNW models on group manifolds 
(this construction is possible for any compact 
or noncompact group) 
\lref\WZW{
J. Wess and B. Zumino, Phys. Lett. B {\bf 37}, 95 (1971)\semi
S. P. Novikov, Sov. Math. Dock, {\bf 24}, 222 (1981)\semi
E. Witten, Comm. Math. Phys. {\bf 92}, 455 (1984).  
} \WZW. 
{\it ii)} WZNW models on particular coset manifolds  
(this construction seems only possible if the structure constants with 3 coset indices are invariant with respect to the subgroup, see for example 
\lref\MT{
R.~R.~Metsaev and A.~A.~Tseytlin,
Nucl.\ Phys.\ B {\bf 533}, 109 (1998)
[hep-th/9805028]; 
R.~R.~Metsaev and A.~A.~Tseytlin,
Phys.\ Rev.\ D {\bf 65}, 126004 (2002)
[hep-th/0202109].
} 
\MT
) {\it iii)} 
gauged WZNW models
\lref\Polyakov{
A.~M.~Polyakov and P.~B.~Wiegmann,
Phys.\ Lett.\ B {\bf 131}, 121 (1983)\semi 
A.~M.~Polyakov and P.~B.~Wiegmann,
Phys.\ Lett.\ B {\bf 141}, 223 (1984).
} \Polyakov.  In the latter case 
the gauging is performed by introducing a set 
of gauge fields $A_{z}$ and $A_{\bar z}$ 
coupled to the currents of a subgroup. 
The gauge fixing $A_{\bar z}=0$ 
leads to a second WZNW model with $h$-currents and ghost currents. 
Gauging once again this BRST invariant action yields the sum of 
gauge, $h$ and ghost currents which forms the starting point for the 
construction of the final BRST charge 
\lref\KarabaliDK{
D.~Karabali and H.~J.~Schnitzer,
Nucl.\ Phys.\ B {\bf 329}, 649 (1990).
} \KarabaliDK. 
This BRST charge 
implements the constraints at the level of the Fock space and selects  
the physical subspace of the theory. 

In the present paper we present a new 
method to obtain interesting nonlinear sigma models by imposing (gauging) the constraints 
related to coset generators and not those of a subgroup. This is 
inspired by recent developments in superstring theory.  
To construct a quantized superstring one may begin 
with the set of second class constraints $d_{z\a}$ (the 
conjugate momenta of the fermionic coordinates $\t^{\a}$) 
as starting point. A BRST charge $Q = \oint \l^\a d_{z\a}$ is constructed and it 
is made nilpotent by imposing suitable quadratic constraints on the 16 complex commuting ghosts $\l^\a$.  The operators $d_{z\a}$ generate a particular non-semisimiple superalgebra, with further generators $\Pi_{zm}$ and $\partial_z \t^\a$ which form a sub-superalgebra. 
In one (noncovariant) approach one imposes the constraint 
$Q | {\rm phys} \rangle =0$ on the physical states.  
\lref\berko{
N.~Berkovits,  
JHEP {\bf 0004}, 018 (2000); 
N.~Berkovits,  
Int.\ J.\ Mod.\ Phys.\ A {\bf 16}, 801 (2001)  
[hep-th/0008145]; N.~Berkovits,
{\it ICTP lectures on covariant quantization of the superstring,}
[hep-th/0209059].
}
\berko. So in this approach one 
gauges $d_{z\a}$. However, since the ghost fields are constrained, gauging 
$d_{z\a}$ does not imply that all corresponding conjugate variables $\t^{\a}$ 
are removed by the cohomology. Rather, the dependence of the vertex 
operators on $\t^{\a}$ is only restricted by the field equations. 

In another (covariant) approach one introduces by hand a ghost doublet 
$(b,c_{z})$ in order to obtain a nilpotent BRST charge, and 
one imposes all three constraints, but then one restricts the carrier space by imposing, a ``grading condition" in 
order that not all $x^{m}$ and $\t^{\a}$ are removed from the 
cohomology
\lref\grassi{
P.~A.~Grassi, G.~Policastro, M.~Porrati and P.~van Nieuwenhuizen,  
JHEP {\bf 10} (2002) 054, 
[hep-th/0112162]; P.~A.~Grassi, G.~Policastro, and P.~van Nieuwenhuizen,  
JHEP {\bf 11} (2002) 004, 
[hep-th/0202123].} 
\lref\grassiWZW{
P.A. Grassi, G. Policastro and P. van Nieuwenhuizen, 
Nucl. Phys. B {\bf 676} 43, 2003.
} \grassi. 
With this grading condition one obtains the same spectrum as from the noncovariant approach, so one has undone the gauging of $\Pi_{zm}$ and $\partial_z \t$ in some sense.  

Although imposing the grading condition by hand yields to correct cohomology, 
we have suspected for a long time that there exists another charge whose 
vanishing on physical states achieves the same purpose. As a first 
step in this direction we have recently removed the ghost pair $(b,c_{z})$ 
(and also another ghost pair $(\omega, \eta_{z})$ which we also 
introduced by hand to have vanishing central charge) by ``gauging'' 
the WZNW model based on superalgebra of $d_{z\a}, \Pi_{zm}$ and 
$\p_{z} \t^{\a}$ \grassiWZW. The procedure of gauging leads to an extra set of currents $d^{(h)}_{z\a}, \Pi^{(h)}_{zm}$ and $\p_{z}\t^{(h)\a}$ with opposite 
sign for the double poles, and as a consequence it has automatically a nilpotent 
BRST charge and a vanishing central charge. So in this approach there 
are no longer any ghosts added by hand. The second step is then to find the 
charge which takes over the role of the grading. 

In this article we give a general construction of such a charge. It turns out 
to be a second BRST charge which anticommutes with the usual BRST 
charge and which arises naturally in the process of 
``gauging coset generators".  We shall  consider a general simple Lie algebra, instead of the nonsemisimple Lie algebra which appears in the string model. The two main ideas on which 
our approach is based are, on the one hand, the structure of Lie 
(or affine Lie) algebras on the Cartan-Weyl basis, and, on the other hand, 
the BRST approach to second class constraints. 

It is well-known how to gauge a subgroup $H$ of a Lie group $G$: one decomposes the generators into coset generators $K_\a$ and subgroup generators $H_i$ 
and one constructs the BRST charge $Q=c^iH_i+ {1\over 2} 
b_if^i_{\ jk}c^kc^j$ where $c^i,b_i$ are
the ghosts and the antighosts, respectively, 
and $f^i_{~ jk}$ the structure constants of $H$. The $H_{i}$ are then 
first class constraints which annihilate physical states: 
$H_{i} | phys \rangle =0$ (which becomes $Q |phys \rangle =0$ in the 
BRST approach). Often an explicit representation
of $K_{\a}$ and $H_{i}$ in terms of differential operators or 
conformal field theory is given. 

The main difference between gauging a subgroup and 
gauging a coset boils down to the fact that the generators of the 
subgroup form a closed algebra of constraints 
whereas the generators of the coset do not. From a Hamiltonian point of view 
this means that the constraints imposed by $K_{\a}$ are second class constraints. Namely, if $K_{\a}$ is a first class constraint, it 
annihilates physical states $K_{\a} |phys \rangle =0$, but then 
also $[K_{\a}, K_{\b}] |phys \rangle =0$. The set of generators 
$[K_{\a}, K_{\b} ]$ in general closes on the 
generators $H_{i}$ (and only $H_{i}$ if one has a symmetric coset 
decomposition). If $H_{i}$ is 
not to be a constraint, $H_{i} | phys \rangle$ should not vanish for 
all physical states, but then $[K_{\a}, K_{\b}] |phys \rangle$ 
should be non-vanishing. Thus gauging coset generators leads to second class 
constraints. 
Using the Dirac brackets to implement the constraints and 
eliminating 
the variables associated with the second class constraints, in general 
reduces the isometry group of target space and looses the manifest 
symmetry of the theory. 
(In the case of superstrings, separating the second class fermionic 
constraints from the first class constraints, one necessarily breaks the 
super-Poincar\'e covariance of the Green-Schwarz sigma model.) 
There is, however, a way to preserve all symmetries, namely by using 
BRST methods. This paper presents a covariant BRST approach 
to the gauging of coset generators.

The basic idea is to start 
with a BRST charge $Q_{K,0}= \xi^{\a} K_{\a}$ instead of the 
coset generators $K_{\a}$ by themselves. The nilpotency of 
$Q_{K,0}$ requires that 
$\xi^{\a} \xi^{\b} [K_{\a}, K_{\b}] $ vanishes, and this can be achieved by imposing suitable constraints on the ghosts $\xi^{\a}$. 
Next we relax these constraints, which requires the introduction of further 
ghosts, but in order that the cohomology does not depend on the extra ghosts, we need a second BRST charge. 
As we already mentioned, 
this is the main idea upon which 
a new approach to the quantization of the superstring is based. 
In that particular example, $K_{\a} = d_{\a}$ (cf. \berko)
are the conjugate momenta of the 
spacetime fermionic coordinates $\t^{\a}$, but in this article we want to 
develop a general formalism applicable to any physical system. 

The content of this paper is as follows: in sec. 2, we 
derive the constraints on the ghost fields and we define the constrained cohomology. 
In sec. 3 we remove the constraints by adding new ghost fields while keeping 
the extended BRST charge $Q_{K}$ nilpotent. 
At the same time we introduce a new nilpotent BRST operator $Q_{C}$  
whose role is to remove the new ghosts from 
physical observables. Without $Q_{C}$ 
there is no cohomology 
for $Q_{K}$ but with $Q_{C}$ the pair $(Q_{K}, Q_{C})$ 
leads to nontrivial ``relative cohomology''. 
We mention that the idea of using a second BRST 
charge has been proved useful in string theory 
\lref\bvw{
N.~Berkovits and C.~Vafa,
Mod.\ Phys.\ Lett.\ A {\bf 9}, 653 (1994)
[hep-th/9310170].
N.~Berkovits and C.~Vafa,
Nucl.\ Phys.\ B {\bf 433}, 123 (1995)
[hep-th/9407190]\semi
N.~Berkovits, C.~Vafa and E.~Witten,
JHEP {\bf 9903}, 018 (1999)
[hep-th/9902098].
} \bvw, in a 
6 dimensional supersymmetric formulation of superstrings on a 
Calabi-Yau manifold
\lref\BerkovitsDU{
N.~Berkovits,
Nucl.\ Phys.\ B {\bf 565}, 333 (2000)
[hep-th/9908041].} \BerkovitsDU, in topological field 
theory \stora, in string field theory 
\lref\GaiottoYB{
D.~Gaiotto and L.~Rastelli,
hep-th/0312196.
} \GaiottoYB, and very recently in a string-inspired formulation 
of harmonic superspace 
\lref\GrassiXC{
P.~A.~Grassi and P.~van Nieuwenhuizen, hep-th/0402189.
} \GrassiXC.
In section 4 we find as an unexpected bonus the solution to a problem 
that has kept us thinking for a long time. In our work on the 
covariant quantum superstring we found the need to introduce a new 
quantum number for the ghosts, called grading, whose role was 
to restrict the vertex operators such that nontrivial cohomology resulted. 
We now propose that this grading has a group theoretical meaning 
which is intimately linked to the second BRST charge.  
In sec. 5, we construct the currents of a 
conformal field theory obtained by gauging the coset generators of 
$SU(N)$, and extend the discussion 
from the affine Lie algebra to the Virasoro algebra and beyond. 
In sec. 6 we mention possible applications of the formalism developed in this 
article. In an appendix we illustrate our method by explicitly 
working through the example of $SU(3)$. 


\newsec{Gauging the coset: the first BRST charge $Q_{K}$}

Consider a simple Lie algebra decomposed into 
the Cartan-Weyl basis of raising operators $E_{\a}$ associated 
with positive roots, lowering operators
$E_{-\a}$ associated with negative roots, 
and Cartan generators $H_i$
$$
[E_\a,E_{-\a}]=\a^i H_i, ~~~~~ 
[H_i, E_{\pm\a}]=\pm\a_i E_{\pm\a}, ~~~~~
$$
\eqn\uno{
\left[ E_{\a},E_\b \right]= N_{\a,\b} E_{\a+\b} 
~~{\rm if} ~~ \a+\b \neq 0\,, ~~~~~ [H_{i},H_{j}]=0\,.
}
We choose the phase convention $N_{\a,\b} = - N_{-\a,-\b}$, and 
$E_{-\a} = (E_{\a})^{\dagger}$. The index $i$ of $\a_{i}$ 
has been lowered with the Killing-Cartan metric according to 
$g_{ij} \a^{j} = g_{\a,-\a} \a_{i}$. (Often one normalizes the $E_{\a}$ such that 
$g_{\a, -\a} =1$, but we shall not  require this).
\foot{For the 
super-Poincar\'e group in 10 dimensions generated by $Q_{\a}$ and $P_{m}$, 
the Killing-Cartan metric is zero. One can 
modify the Lie algebra by adding a new central charge, 
and in this way one obtains 
a non-degenerate metric \lref\grr{
M.B. Green, 
Phys. Lett. B {\bf 223} (1989) 157; 
W.~Siegel, 
Nucl. Phys. B {\bf 263} (1986) 93;   
W.~Siegel, 
Phys. Rev. D {\bf 50} (1994) 2799.  
} \grr.     We propose that in such cases one uses this metric 
to lower the index of $\a^{i}$. 
}
We begin with the BRST charge
\eqn\due{
Q_{K,0}=\sum_{\a \in \Delta}\xi^{\a} \, E_{\a} - {1\over 2}
\sum_{\a,\b \in \Delta}^{\a+\b \neq 0} \b_{\a+\b} \, N_{\a,\b} \, \xi^\a\xi^\b
}
where the sum is over all roots (we denote by $\Delta$ the set of all roots and 
$\Delta_{+}$ the set of positive roots). The $\xi^{\a}$ are anticommuting 
ghosts and $\beta_{\a}$ are the corresponding antighosts. 
One may view this as an expression in quantum mechanics with brackets $\{\xi^\a,\b_\b\}=\d^\a_{\ \b}$
or an expression of the holomorphic sector in string theory 
with operator product expansion (OPE) $\xi^\a(z)\b_\b(w)\sim
\d^\a_{\ \b} {1 \over z-w}$. (In the latter case one should add an integration 
$\oint dz$ to the definition of $Q_{K,0}$).
Nilpotency of $Q_{K,0}$ requires the following constraints\foot{In the case of superstrings 
the generators of the coset are represented by 
covariant derivatives $d_{z\a}$ and 
the proper maximal subgroup is generated by the translations 
$\Pi^{m}_{z}$ and the 
fermionic translations $\partial_{z}\theta^{\a}$. The constraints (2.3) 
correspond to 
$\l^{\a} \g^{m}_{\a\b} \l^{\b}=0$ for the commuting spinors $\l^{\a}$, and 
such $\l^{\a}$ are called pure spinors.   
}
\eqn\tre{
\sum_{\rm \a \in \Delta_{+}} \a^i \xi^\a \xi^{-\a}=0 \,,
}
where the index $i$ labels the Cartan generators. 
For example, if the group is  $SU(2)$, the two 
ghosts $\xi^{\pm}$ correspond 
to the raising and lowering 
generators $E_{\pm}$ and the constraint \tre\ becomes 
$\xi^+\xi^- = 0$ of which $\xi^-=\xi^+$ is a solution.\foot{Notice that for Grassmann variables one 
has $\xi^{+} \equiv 
\delta(\xi^{+})$, the Dirac delta function with respect to the Berezin 
integration, $\int d\xi^{+} \delta(\xi^{+}) f(\xi^{+}) = \int d\xi^{+} \xi^{+} f(\xi^{+}) = f(0)$. For $SU(3)$, one has the constraints 
$\xi^{1} \xi^{-1} + {1\over 2} \xi^{2} \xi^{-2} + {1\over 2} \xi^{3} \xi^{-3} =0$ 
and $\xi^{2} \xi^{-2} - \xi^{3} \xi^{-3} =0$, which can be solved by setting $\xi^{\pm 1}= \pm
\xi^{ \pm 2} = \pm \xi^{\pm 3}$ or by the minimal solution $\xi^{\a} =\xi^{-\a}$ for 
all $\a$.} 
There are thus solutions in general, but they break the $H$-invariance 
(in the case of the 
ten-dimensional superstring any solution of the constraints $\l \g^{m} \l =0$ breaks the  manifest Lorentz covariance). The 
constraints in \tre\ clearly 
commute with each other and are invariant under the BRST 
transformations generated by 
\due\ (see below). 
Hence they are first class constraints on the ghost fields. 
They generate gauge transformations 
on the antighosts $\beta_{\a}$ 
\eqn\quattro{
\delta_{\e} \beta_{\a} = 
\Big[ \e_{i} 
\sum_{\rm \b \in \Delta_{+}} \b^i \xi^\b \xi^{-\b}, \beta_{\a} \Big] = 
- \e_{i} \a^{i} \xi^{-\a}
}
where $\e^{i}$ are infinitesimal local parameters (one for 
each generator of the Cartan subalgebra). 

In order that the constraints \tre\ are compatible with 
the BRST symmetry, they should be invariant under it. 
One can check by using the 
Jacobi identity $[[E_{\a}, E_{\b}], E_{-\a-\b}] + [[E_{\b}, E_{-\a-\b}], E_{\a}] +
[[E_{-\a-\b}, E_{\a}], E_{\b}] =0$ that this is the case
\eqn\cinque{
[Q_{K,0}, \delta_{\e}] =  \Big[ \sum_{\a \in \Delta}
\xi^{\a} \, E_{\a} - {1\over 2} 
\sum^{\a +\b \neq 0}_{\a,\b \in \Delta} 
\b_{\a+\b} \, N_{\a,\b}\xi^\a\xi^\b, 
\sum_{i} \e_{i} \sum_{\rm \a \in \Delta_{+}} \a^i \xi^\a \xi^{-\a} \Big] =0\,.
} 

The space ${\cal M}$ on which 
the Lie  algebra acts is the group manifold 
${\cal G}$ parametrized by a set of coordinates; in that case 
the generators $E_{\a}$ can be 
represented by differential operators
 $D_{\a}$ 
acting on functions defined on ${\cal G}$. We extend ${\cal M}$ to the 
space $\widehat{\cal M}$ which contains the ghosts $\xi^{\a}$. 
Then, we impose the constraints \tre\ to define the reduced 
functional space $\widehat{\cal M}'$ 
on which we compute the cohomology $H(Q_{K,0}, \widehat{\cal M}')$. 

The space $\widehat{\cal M}'$ decomposes into subspaces 
$\widehat{\cal M}'_{(n)}$
with ghost number $n$. Consider the sector with ghost 
number one, containing the following functions 
\eqn\tredici{
\Phi^{(1)} = \sum_{\a \in \Delta}\xi^\a A_\a({\cal M}).
}
where $A_\a({\cal M})$ is a function on the group manifold ${\cal M}$.  
Acting with $Q_{K,0}$ in \due\ on $\Phi^{(1)}$ while imposing
the constraint \tre\ leads to restrictions on the fields $A_{\a}$
\eqn\quattordici{
[Q_{K,0},\Phi^{(1)}]={1\over 2} \sum_{\a,\b \in \Delta}
\xi^\a\xi^{\b} \Big( D_\a A_{\b}-D_{\b}A_\a  - 
N_{\a,\b} A_{\a+\b}\Big) = 0\,. 
}
The general solution of this equation is 
\eqn\quindici{
F_{[\a,\b]} \equiv 
\Big(
D_\a A_{\b} - D_{\b}A_\a - N_{\a,\b} A_{\a+\b}
\Big) = \d_{\a+\b,0} \sum_{i} \a^iW_i\,.
}
The left hand side can be viewed as the curvature $F_{[\a,\b]}$ 
of the group manifold along all roots. 
We have $F_{\a,-\a} = \sum_{i} \a^{i} W_{i}$ 
for each root $\a \in \Delta_{+}$.\foot{A rather simple example is the 
$SU(2)$ case. In that case $\Phi^{(1)} = \xi^{+} A_{+} + \xi^{-} A_{-}$ and 
$\{Q _{K,0},\Phi^{(1)}\} = \xi^{+} \xi^{-} (D_{+} A_{-} - D_{-} A_{+}) = 0$ 
due to the constraints $\xi^{+} \xi^{-} =0$. This implies that any function 
$A_{\pm}$ such that $A_{\pm} \neq D_{\pm} \Omega$ 
(where $\Omega \in {\cal M}_{(0)}$) belongs to the cohomology 
$H^{(1)}(Q_{K,0}| \widehat{\cal M}')$. In addition, it is obvious 
that $\widehat{\cal M}'_{(n)} = \emptyset$ for $n \geq 2$. 
Finally, we have $\Phi^{(0)} = A$ where $A \in {\cal M}$. Acting 
with $Q_{K,0} $ one finds $\xi^{+} D_{+} A + \xi^{-} D_{+}A =0$ 
and using the solution $\xi^{+} = \xi^{-}$, it follows that 
$(D_{+} -  D_{-}) A =0$ which has nontrivial solutions. 
So both $H^{(0)}(Q_{K,0}|\widehat{\cal M}')$ and 
$H^{(1)}(Q|\widehat{\cal M}')$ are not empty.} 
Because the curvatures are  
non-vanishing when $\a = -\b$, the fields $A_{\a}$ are nontrivial 
(not pure gauge). By acting with $E_{\gamma} + A_{\gamma}$ on $F_{\a\b}$ and summing over all cyclic permutations of $\a,\b,\g$, the Bianchi identities
yield constraints on $A_{\a}$ and $W^{i}$, but we do not analyze 
these issues here further. 
One could completely gauge the group ${\cal G}$ 
by adding the Cartan generators multiplied by 
the corresponding ghosts to the 
BRST charge (see below). In that case, all curvatures 
vanish, implying that there are no propagating physical degrees of freedom.  

There are solutions of \quindici\ which are given by the BRST exact elements 
of $\widehat{\cal M}'$. For example at ghost number one we may pick a 
function $\Omega \in \widehat{\cal M}'_{(0)}$ with ghost number zero and 
acting with $Q_{K,0}$ on it, one obtains the gauge transformation 
\eqn\diciasette{
\delta \Phi^{(1)} = \{Q_{K,0}, \Omega\} = 
\sum_{\a} \xi^{\a} D_{\a} \Omega\,.
}
This gives the gauge transformations 
$\delta A_{\a} = D_{\a} \Omega$. By inserting 
this gauge transformation into \quindici\ one gets 
$\delta W_{i} = D_{i} \Omega$ where $D_{i}$ is the differential 
operator acting 
on the Cartan coordinates of the group manifold. The 
solution of (2.8) is then given by $A_{\a} = D_{\a} \Omega$ and 
$W_{i} = D_{i} \Omega$. By assuming that 
the group manifold is independent of those coordinates we obtain 
that $W_{i}$ are invariant under the gauge transformation. 

For later use, we note that 
the Casimir operator $C_{2}$ is given by 
\eqn\diciotto{
C_{2} = \sum_{\a \in \D_{+}} 
g^{\a, -\a}
(E_{\a} E_{-\a} + E_{-\a} E_{\a})
+ \sum_{i,j} g^{i j} H_{i} H_{j}
}
and it is BRST invariant, $[Q_{K,0}, C_{2}]=0$.  


In the next section we relax the constraints, but before moving on 
we would like to point out that the approach to constrained 
systems of this section is already a generalization of the pure spinor 
formalism \berko\ for the superstring 
to a wider class of models. From this point of view 
Berkovits' pure spinor formulation corresponds to the gauging of the 
coset currents of a particular WZNW model \grassiWZW. 


\newsec{No constraints: the second BRST charge $Q_{C}$}

Working with constrained fields is 
not very practical and therefore it is desirable  to remove the 
constraints. The most straigthforward way to remove them is to 
implement these constraints at the level  of the BRST cohomology, by adding 
new ghosts which are Lagrange multipliers. 
By requiring nilpotency of the 
BRST charge, further terms in the BRST charge $Q_{K}$ can be 
determined. However, 
this procedure renders the cohomology empty (except for a few 
non-propagating degrees of freedom at zero momentum 
\ganor). Therefore, we 
develop a method which does recover the correct cohomology. 
First we construct the full BRST charge by introducing new ghosts. Then 
we construct a second BRST charge $Q_{C}$ 
whose role is to remove the new 
ghosts and to yield the same nontrivial cohomology 
as we started with.\foot{A similar analysis for 
the superparticle has been pursued in  
 \lref\chesterman{
M.~Chesterman, 
JHEP {\bf 0402}, 011 (2004) 
[hep-th/0212261].
} \chesterman.}

To relax the constraints \tre\ we add two further terms 
to the BRST charge $Q_{K,0}$
\eqn\quattro{
Q_{K,-1}= - \sum_{\a \in \Delta_{+}} 
\bar{\eta}_i \a^i\xi^\a\xi^{-\a}\,, 
~~~~~~~~
Q'_{K,1}= \sum_{i} \eta^i \, H_i\,, 
}
The anticommuting ghosts $\eta^{i}$ and $\bar\eta_{j}$ satisfy the bracket 
$\{\eta^{i}, \bar\eta_{j}\} = \delta^{i}_{j}$ or $\eta^{i}(z) \bar\eta_{j}(w) \sim \delta^{i}_{j}
(z-w)^{-1}$. At this stage, 
$(Q_{K,-1}+Q_{K,0}+Q'_{K,1})^2$
contains only terms of the form $ \sum_{i\a} \a_i \eta^i \xi^\a E_\a $, and
they can be canceled by adding the usual three ghost term
\eqn\cinque{
Q''_{K,1}=  \sum_{i, \a \in \Delta} 
\a_{i} \, \eta^{i} \, \beta_{\a}  \xi^{\a} \,. 
}
The BRST charge can be 
decomposed into terms $Q_{K,n}$ with different grading $n$
if one assigns the following grading to the ghosts and 
antighosts
\eqn\grading{
gr(\xi^{\a})=0,~~~~~~~
gr(\eta^i)=1, ~~~~~
gr(\b_{\a})=0~~~~~
gr(\bar{\eta}_i)=-1\,.
}
The nilpotency of the BRST charge $Q_{K}$ and the existence 
of this grading lead to a filtration of the nilpotency relations 
\eqn\grade{
Q_{K,-1}^{2} =0\,, ~~~~~ 
\{Q_{K,-1}, Q_{K,0}\}=0\,, ~~~~~~
\{Q_{K,0},Q_{K,0}\} + 2 \, \{Q_{K,1}, Q_{K,-1}\}=0\,, ~~~~
}
$$
\{Q_{K,0}, Q_{K,1}\} =0\,, ~~~~ 
Q_{K,1}^{2} =0\,.
$$
The second equation 
implies the invariance of the constraints \tre. The third equation tells 
us that the charge $Q_{K,0}$ is nilpotent up to the constraints \tre. 
The fourth relation holds since it is proportional to the sum 
of $(\a+\b)_{i}$, $-\a_{i}$ and $-\b_{i}$. 
The last equation expresses the simple fact that the Cartan generators are 
abelian $(\eta^{i} \eta^{j} H_{i} H_{j} =0)$. 

The new BRST operator $Q_{K}$ has trivial cohomology. 
(We will demonstrate this later in a model of conformal field theory where 
the BRST operator can be obtained from a $G/G$ gauged WZNW model.) 
This can be understood  as follows: the BRST charge 
$Q_{K}$ contains the operators $E_{\a}, E_{-\a}$ and $H_{i}$ for the 
roots and Cartan subalgebra. Therefore, the BRST closed 
functions of $\widehat{\cal M}_{(1)}$ ({\it i.e.} the unconstrained space 
with all ghosts) are given by 
\eqn\XX{
\Phi^{(1)} = \sum_{\a } 
\xi^{\a} A_{\a} + \sum_{i} \eta^{i} A_{i}\,.
}
By definining the curvatures of the fields of 
$A_{\a}$ and $A_{i}$ as usual
$$
F_{[\a,\b]}= D_{[\a} A_{\b]} - N_{\a,\b} A_{\a+\b} ~~{\rm if} ~~\a+\b \neq 0 
~~~~~~
F_{\a, -\a} = D_{[\a} A_{-\a]} - \a^{i} A_{i} \,, 
$$
\eqn\xx{
F_{\pm \a,i} = D_{i} A_{\pm\a} - D_{\pm\a} A_{i} \mp \a_{i} A_{\a} \,, ~~~~~~~
F_{i,j} = D_{[i} A_{j]}\,, 
}
the equation $\{Q_{K}, \Phi^{(1)} \} =0$ 
implies that all curvatures vanish. Therefore, we 
can start solving this system by observing that $A_{i} = D_{i} \Omega$ 
solves locally $F_{i,j} =0$, and inserting this result in 
$F_{\pm\a,i} =0$, one gets $A_{\pm\a} =D_{\pm\a} \Omega$. So, the general 
solution of \xx\ are pure gauge connections which corresponds to 
BRST exact $\Phi^{(1)} = \{Q_{K}, \Omega\}$. The 
same results hold for other ghost numbers. 
In order not to remove the nontrivial cohomological classes 
of $\widehat{\cal M}'$, we have to establish a new definition of physical observables. To this purpose
we introduce a second BRST 
charge $Q_{C}$. This requires to extend again the set of ghost fields. 

The additional ghosts have the role to remove the 
ghosts $\eta^{i}$ and $\bar\eta_{i}$ from the cohomology. 
We add a quartet (two new doublets) of fields. The first 
doublet contains a pair of 
anticommuting fields $\eta^{\prime i}$ and $\bar\eta'_{i}$ 
with brackets $\{\eta^{\prime i}, \bar\eta_{j}'\} = \delta^{i}_{j}$ and 
with the same quantum number as $\eta^{i}$ and $\bar\eta_{i}$, 
where the index $i$ runs over the Cartan subalgebra. In addition, 
we introduce two commuting fields $\phi^{i}$ and $\bar\phi_{i}$ with 
brackets $[\phi^{i}, \bar\phi_{j}]  = \delta^{i}_{j}$. 
The new ghosts have the following ghost and grading numbers, respectively 
\eqn\xxi{
\eta^{\prime i} ~~~ (1,1)\,, ~~~~
\bar\eta'_{i} ~~~ (-1,-1)\,, ~~~~
\phi^{i} ~~~ (0,0)\,, ~~~~
\bar\phi_{i} ~~~ (0,0)\,. ~~~~
}

We follow now the procedure of \stora. The 
$(\phi, \bar\phi)$ form with the $(\eta, \bar\eta)$ a quartet of 
$Q_{C}$,  but the same $(\phi, \bar\phi)$ form with $(\eta', \bar\eta')$ 
another quartet of $Q_{K}$. In this way, we remove all six ghosts 
$\eta, \bar\eta, \eta', \bar\eta', \phi, \bar\phi$ from the cohomology. 
The new BRST charge $Q_{C}$ is given by
\eqn\XXI{
Q_{C} = \sum_{\a \in \Delta_{+}, i} \bar\eta'_{i} \a^{i} \xi^{\a} \xi^{-\a} + 
\sum_{i} \bar\phi_{i} \eta^{i} \,. 
}
It is obviously nilpotent. 
The second term removes $\bar\eta^{i}$ (and its conjugate $\eta_{i}$ 
can be set to zero) from the space 
$\widehat{\cal M}$.\foot{Since $Q_{C} \bar\eta_{i} = \bar\phi_{i}$ 
and $Q_{C} \phi^{i} = - \eta^{i}$, any function $F(\eta, \phi, \eta', \dots)$ 
which is annihilated by $Q_{C}$ is independent 
of $\phi$ and $\eta$ 
up to terms which 
are $Q_{C}$ exact.  In fact, there is an homotopy operator $K_{C}$, satisfying $\{K_{C}, Q_{C} \} = N_{\phi, \bar\phi} + N_{\eta,\bar\eta}$ and 
given by $K_{C} = \phi^{i} \bar\eta_{i}$. Then any state with  
$\phi$ and $\eta$ dependence which is $Q_{C}$-closed  
is also $Q_{C}$-exact. Similarly, $Q_{K} \bar\eta'_{i} = \bar\phi_{i}$ and 
$Q_{K} \phi^{i} = - \eta^{\prime i}$, and $Q_{K}$ removes the $\phi_{i}$ dependence while terms depending on $\eta'_{i}$ are $Q_{K}$-exact. 
}
The first term is needed in order that the second BRST charge commutes 
with the original charge $Q_{K}$. However, we also 
have to add a new piece to $Q_{K}$ 
\eqn\XXII{
Q_{K} \rightarrow Q_{K} + \sum_{i} \bar\phi_{i} \eta^{\prime i}\,.
}
This extended $Q_{K}$ is clearly also nilpotent. 

The new term in $Q_{K}$ removes 
the variables $\eta', \bar\eta', \phi$ and $\bar\phi$ 
from the cohomology  of $Q_{K}$. Although we keep all ghosts in our 
covariant approach, note that if we would remove all of them, we would 
have to impose by hand the original constraints in \tre. The 
argument is the same as in the case of the $A_{0}=0$ gauge in QED, where 
one has to impose by hand its missing field equation, the Gauss constraint. 
Notice that the second BRST charge has only terms with grading number 
$-1$ and $+1$. 

The two BRST charges anticommute 
\eqn\XXIII{
Q^{2}_{K} = 0\,, ~~~~~~~~~ 
\{Q_{K}, Q_{C}\} = 0\,, ~~~~~~~~ 
Q_{C}^{2} = 0 \,.
} 
To prove that the terms with $N_{\a,\b}$ cancel in $\{Q_{K}, Q_{C}\}$ one 
may use the Jacobi identities with $E_{\a}, E_{\b}$ and $E_{-\a-\b}$. The 
terms proportional to $\bar\eta' \eta$ cancel because they occur in pairs 
with opposite signs. 
In addition, it is easy to see that the first BRST charge 
$Q_{K}$ can be written in the following way
\eqn\XXV{
Q_{K} = \sum_{\a} \xi^{\a} E_{\a}  - {1 \over 2}  
\sum_{\a,\b} \b_{\a+\b} N_{\a,\b} \xi^{\a} \xi^{\b} +
\left[Q_{C} , \sum_{i} 
\Big[ \bar \eta_{i} \eta^{\prime i} - \phi^{i} ( H_{i} + \sum_{\a} 
\a_{i} \beta_{\a} \xi^{\a}) \Big]\right]\,,
}
which shows that the additional terms are indeed trivial 
with respect to the second BRST charge. 
Furthermore, $Q_{C}$ is $Q_{K}$-exact, namely
\eqn\xxv{
Q_{C} = \Big[ Q_{K}, \bar\eta'_{i} \eta^{i} \Big] \,.
}

The most general vertex with ghost number one 
can be written as follows 
\eqn\XXVI{
\Phi^{(1)} = \sum_{\a} \xi^{\a} A_{\a} + \Big\{ Q_{C}, \sum_{i} 
\phi^{i} W_{i} \Big\}\,.
}
The second term is $Q_{C}$-trivial and it is needed in order that 
$\Phi^{(1)}$ is in the cohomology of $Q_{K}$. 
Physical states are identified with 
$H(Q_{K} | H(Q_{C}, \widehat{\cal M}))$: they are $Q_{C}$-closed, and 
$Q_{K}$-closed modulo $Q_{C}$-exact terms.
To work this out in more detail note that the $Q_{K}$ cohomology can be 
written as follows
\eqn\XXVII{
\{ Q_{K}, \Phi^{(1)}\}  = \Big\{ Q_{K,0} + 
\Big[ Q_{C}, X \Big] , \sum_{\a} \xi^{\a} A_{\a} 
+ \Big[ Q_{C}, \sum_{i} \phi^{i} W_{i} \Big]\Big\} =
}
$$
\Big\{ Q_{K,0} , \sum_{\a} \xi^{\a} A_{\a} 
\Big\} + \Big\{ Q_{C} , \Big[ X, \sum_{\a} \xi^{\a} A_{\a} \Big] - 
\Big[Q_{K,0}, 
\sum_{i} \phi^{i} W_{i} \Big] + 
\Big[ X, \Big[ Q_{C}, \sum_{i} \phi^{i} W_{i} \Big]\Big]
\}=
$$
$$
\Big\{ Q_{K,0} , \sum_{\a} \xi^{\a} A_{\a} \Big\} + \{Q_{C}, Z \}\,.
$$
where $X$ is the $Q_{C}$-exact term in \XXV\ and $Z$ is the 
$Q_{C}$-exact term in \XXVI. We used 
$\{Q_{C}, \sum_{\a} \xi^{\a} A_{\a} \} =0$. 
This shows that we have obtained the correct 
cohomology: the $Q_{K}$-closed vertex operators constructed from 
all ghosts given modulo 
the $Q_{C}$-exact terms coincide with the vertex operators which are 
constructed from $\xi^{\a}$, and which are $Q_{K,0}$-closed 
modulo the constraints in \tre. 


\newsec{An interpretation of the grading}

In our work on the covariant quantization of the superstring we were forced 
to exclude certain terms from the massless vertex operators in order to 
obtain a nontrivial cohomology. We achieved this by assigning a grading 
to the various ghosts which appear in our work, and requiring that vertex 
operators contain only terms with nonnegative overall grading. The 
deep meaning of this grading has eluded us up till now although we have 
shown that it is related to homological perturbation theory
\lref\GrassiCM{
P.~A.~Grassi, G.~Policastro and P.~van Nieuwenhuizen,
Class.\ Quant.\ Grav.\  {\bf 20}, S395 (2003)
[arXiv:hep-th/0302147].
} \refs{\grassi,\GrassiCM}. 
We now present an interpretation of this grading condition. 

In the previous section, we used a second BRST charge to select 
the physical subspace. This suggests that there exists another 
quantum number for the 
ghosts and antighosts besides the ghost number. The is the quantum number 
given in \grading\ and \xxi. However, the notion of a grading was extracted from 
the BRST approach, and 
we should explore whether the grading has a meaning 
independently of this construction, in particular whether there is a property of  
Lie algebras which leads to the concept of this grading. 

Let us return to 
the Lie algebra on the Cartan-Weyl basis 
\eqn\CW{
[H_{i}, H_{j}] =0\,, ~~~~~~~
[E_\a,E_{-\a}]=\a^i H_i,
}
$$ 
[H_{i}, E_{\pm\a}]=\pm\a_i E_{\pm\a}, ~~~~~
\left[ E_{\a},E_\b \right]= N_{\a,\b} E_{\a+\b}\,.
$$
 These commutators are preserved by the following 
 transformations
 \eqn\CWII{
 H_{i} \rightarrow \l\, H_{i}\,, ~~~~~~ E_{\a} \rightarrow E_{\a}
 ~~~~~~ \a_{i} \rightarrow \l \, \a_{i}
 }
where $\l \neq 0$.  As a consquence $\a^{i} \rightarrow \l^{-1} \a^{i}$. 
We identify the grading with the power of $\l$ in this automorphism. 
Thus we assign a grading $+1$ to each Cartan generator. 
The transformation rule for the roots $\a_{i}$ (for each component 
we assume the same dilatation) is a consequence of the dilatation of the 
Cartan generators. 

By viewing $E_{\a}$ and $H_{i}$ as constraints on the 
physical states: $E_{\a} |\psi \rangle =0$ 
and $H_{i} |\psi \rangle =0$ the contraction $\l \rightarrow 0$ 
reduces the set of the constraints to those implemented by 
the lowering and raising operators $E_{\a}$ with $\a \in \Delta$, 
but they become second class constraints. In fact the r.h.s. of 
$ [E_\a,E_{-\a}]=\a^i H_i$ has a finite limiting value. This implies that 
we need to implement a quantization procedure which deals with second 
class constraints. 

Applying these considerations to the superstring, we are led to the following 
gradings: $\l^{\a}$ has grading zero, and $\xi^{m}$ and $\chi_{\a}$ 
have grading one. This is not the grading proposed in 
\grassi, but remarkably, it gives the same suppression of terms in the 
vertex operator and thus the same spectrum. (More 
specifically: the terms $b \l^{\a} \l^{\b}$ and $b \l^{\a} \xi^{m}$ 
which had grading $-4 +1 +1$ and $-4 +1 +2$, respectively, were 
rejected. The new grading rejects only the term $b \l^{\a} \l^{\b}$). 

At the level of the BRST charge (assigning zero grading to 
any ghosts and antighosts) we see that by rescaling the 
Cartan generators (and the roots $\a_{i}$) the terms $Q'_{K,1}$ and 
$Q''_{K,1}$ vanish. However, $Q_{K,-1}$ explodes if 
we do not require that $\sum_{\a \in \Delta} \a^{i} \xi^{\a} \xi^{-\a} =0$, 
namely if we do not require that the constraints \tre\ are satisfied. At this 
point it is clear that we can reabsorb the rescaling \CWII\ into the 
ghost fields by assigning a suitable grading or, equivalently, by 
rescaling them. In this way the BRST charge $Q_{K}$ is 
decomposed into the pieces $Q_{K,-1} + Q_{K,0} + Q_{K,1}$ and 
the second BRST charge $Q_{C}$ into 
$Q_{C} = Q_{C,-1} + Q_{C,1}$.

To conclude, the grading is a property of Lie algebras which 
we transfer to the ghost fields.


\newsec{An example: $SU(N)$ conformal field theory}

We now present an example where the general formalism 
developed in the previous sections is worked out explicitly. The example 
is the conformal field theory associated to the affine Lie algebra of 
currents based on $SU(N)$.  
The generators are represented by 
holomorphic currents $E_{\a}(z), H_{i}(z)$. 
We neglect the anti-holomorphic sector of the theory.
There are double poles in the OPE's of these currents 
 \eqn\aI{
E_\a(z)E_{-\a}(w)\sim {\a^iH_i(w) \over z-w}+ {k \over (z-w)^2}\,,
}
$$
E_{\a}(z) E_{\b}(w) \sim N_{\a,\b}{ E_{\a+\b} \over (z-w)} \,, ~~~~{\rm if} ~~~ 
\a+\b 
\neq 0
$$
$$
H_{i}(z) E_{\a}(w)   \sim {\a_{i} E_{\a}(w) \over (z-w)}\,, ~~~~~
H_{i}(z) H_{j}(w) \sim {k\, g_{ij}  \over 2 (z-w)^{2}}\,.
$$
The value of the constant $k$ is the level of the affine Lie algebra we 
are considering and $g_{ij}$ is the Killing-Cartan metric. In order to ``gauge'' 
the currents $E_{\a}(z)$ associated to the positive and negative roots of the 
algebra (which from a quantum field theory point of view is 
equivalent to imposing the constraints $E_{\a}(z) |\psi \rangle = 0$ on physical states 
$|\psi \rangle$) we have to introduce auxiliary currents $E^{h}_{\a}(z)$ and 
$H^{h}_{i}(z)$ which have the same single poles as 
$E_{\a}(z)$ and $H_{i}(z)$, but 
opposite double poles. (This can be achieved by introducing a new WZNW 
model with level $- 2 N - k$ (where $N$ is the dimension of the gauge group 
$SU(N)$)  
and those fields are associated to the gauge fields needed 
to impose the constraints).\foot{In our earlier work, we used the 
formalism with a ghost pair ($c_z,b$), satisfying 
$c_z(z)b(w)\sim {1 \over z-w}$, to remove the anomaly 
in the BRST charge, but recently 
we have dropped the ghosts ($c_z,b$) in favor of the auxiliary 
$h$-currents \grassiWZW.}

The BRST current $j_{K,0}$ corresponding to 
$Q_{K,0}$ introduced in section 2 is given by
\eqn\aII{
j_{K,0}(z)= \sum_{\a \in \Delta} \xi^\a \Big(E_\a(z)+E_\a^{(h)}(z)\Big) - 
{1\over 2} \sum_{\a, \b \in \Delta } N_{\a,\b} \beta_{\a+\b} \xi^{\a} \xi^{\b}
}
which is nilpotent up to the constraints \tre. The combination of 
currents  
$(E_{\a} + E^{(h)}_{\a} + \sum_{\b} N_{\a, \b} \b_{\a+\b} \xi^{\beta})$ does 
not have double poles and it can be used 
to construct the left-moving sector of the BRST charge $Q_{K,1} = 
\oint dz j_{K,0}(z)$. All single poles of the combination 
$(E_{\a} + E^{(h)}_{\a} + \sum_{\b} N_{\a, \b} \b_{\a+\b} \xi^{\beta})$ are 
cancelled except those proportional to the Cartan generators. 
The constraints \tre\ generate gauge transformations of the antighosts and
therefore any conformal field theory operator should be compatible 
with those transformations. 

The definition of physical states is given as in the previous section 
(cf. eq. (2.7)) in the constrained cohomology of $Q_{K,0}$. 
Using the Sugawara construction, the energy-momentum 
tensor is given by 
\eqn\aIII{
T_{zz} = {1\over 2(k + N)} 
\sum_{\a \in \Delta_{+}} 
{|\a|^{2} \over 2} (E_{+\a}E_{-\a} + E_{-\a} E_{\a})}
$$
~~~~~~~
-{1\over 2(k + N)} 
\sum_{\a \in \Delta_{+}} 
{|\a|^{2} \over 2} (E_{\a}^{(h)} E_{-\a}^{(h)} + E_{-\a}^{(h)} E_{+\a}^{(h)})
$$
$$
+ \sum_{i} (H_{i} H_{i} + H^{(h)}_{i} H^{(h)}_{i}) 
- \sum_{\a \in \Delta} \b_\a\p_z\xi^\a \,.
$$
The last term needs some explanation. The ghost fields $\xi^{\a}$ 
are constrained and, as a consequence, the antighosts transform 
under gauge transformation generated by the constraint \tre. This means 
that the constraint eliminates some of the ghosts $\xi^{\a}$ (for example,
for $SU(2)$ there is only one independent ghost field $\xi^{+} = \xi^{-}$) 
and thus $T_{zz}$ (as well as the Lagrangian) depends only on certain 
combinations of the antighost fields ( for $SU(2)$,  the ghost dependent term 
is given by $\b_{+} \p \xi^{+} + \b_{-} \p \xi^{-} = 
(\b_{+} + \b_{-}) \p \xi^{+} = \tilde\b_{+} \p \xi^{+}$ where  
$\tilde\b_{+} = \b_{+} + \b_{-}$ is the combination gauge invariant under 
$\d_{\e}\b_{-} = \e \xi^{+}$ and 
$\d_{\e}\b_{+} = - \e \xi^{-} = - \e \xi^{+}$). The tensor $T_{zz}$ 
is invariant under BRST transformations, $[Q_{K,1}, T_{zz}] = 0$. 

The total conformal charge of the system is 
\eqn\aIV{
c_{_{SU(N)}} = {k (N^{2}-1) \over k + N} - {(- 2N -k) (N^{2}-1) \over k +N} 
- 2 [N(N -1) - (N-1)] = 4 (N-1) 
}
where the last term is due to the ghosts and antighosts 
associated with the $N(N-1)$ roots minus the 
number of the constraints $N-1$ (see \tre). The factor 
$-2$ comes from the conformal weight $(0,1)$ and 
statistics of the pairs $(\xi^{\a}, \b_{\a})$. The total conformal 
charge is always positive and it does not depend on the level 
of the WZW action. Notice that without the constraints and with the ghosts associated to the Cartan generators, 
the last term in \aIV\ would be $- 2 [N(N -1) + (N-1)] =  - 2 (N^{2} -1)$ and 
it cancels exactly the first two terms in $c_{_{SU(N)}}$. 
This coincides with the topological model $G/G$. (The total central 
charge 
vanishes because $T$ is the energy-momentum tensor  for a twisted superconformal algebra.)  The total central charge is positive (the theory 
is unitary) and it can vanish only if $N=1$ which is a trivial case. 
However, another way to make the central charge vanish (except 
for example by adding suitable ghosts for reparametrizations) is to add a 
fermionic counterpart to the generators $E_{\a}$ and $H_{i}$. So, for 
the superalgebra 
$SU(M|N)$, we have $\{E_{\a}, H_{i}\}$ and 
$\{E'_{\a'}, H'_{i'}\}$ for the subgroup $SU(M) \times SU(N)$. 
In addition we have $2 M \times N$ fermionic generators $Q_{a b'}$ 
where $a = 1, \dots, M$ and $b'=1, \dots N$.  In that case we have to 
decide which coset we need to gauge and therefore we have to 
introduce bosonic ghosts. 
 
Given a BRST current $j_{K,z}$ and the energy-momentum 
tensor $T_{zz}$, there is an additional operator worth the be mentioned: 
the ghost current. In the present case it is given by 
$J^{gh}_{z} = \sum_{\a \in \Delta} \b_{\a} \xi^{\a}$ which is invariant under the gauge 
transformations $\delta_{\e}$ generated by the first class constraints \tre. 
In order to compute the coefficient of the double pole of $J(z) J(w)$ 
 first to solve the constraint \tre, then  one can choose 
a gauge for the antighosts.\foot{In the SU(2) case, one has 
$\xi^{+} = \xi^{-}$ and $J = \beta_{+} \xi^{+} + \beta_{-} \xi^{-} = 
(\beta_{+} + \beta_{-}) \xi^{+} = \tilde\beta_{+} \xi^{+}$. The  
last expression involves only free fields and we can compute 
the coefficient straightforwardly $J(z) J(w) \sim (z-w)^{-2}$.}

The next step is to construct the second BRST 
charge $Q_{C}$ for this conformal field 
theory. As a consequence we have to modify the BRST charge 
$Q_{K}$. As in the previous section we introduce the fields 
$\eta^{i}, \bar\eta_{i}$ to remove the ghost constraints \tre\ and 
to implement the constraints associated to the Cartan generators. 
(Notice that the enlargament of the set of the constraints to the complete 
algebra leads to vanishing cohomology unless an addition constraint
is added.) 
The new BRST current is
\eqn\aVI{
j_{K}(z) = j_{K,0}(z) + \sum_{i, \a \in \Delta_{+}} 
\bar{\eta}_i \a^i \xi^{+\a} \xi^{-\a} +  
\sum_{i} \Big[
\eta^i(H_i+H_i^{(h)}) + 
\sum_{\a \in \Delta} 
\eta^i\a_i \b_{\a}\xi^{\a} \Big]
}
and the new energy-momentum tensor is modified into 
$$
T_{zz} \rightarrow  T_{zz} + \bar{\eta}_i \p \eta^i.
$$
By adding the new fields and by modifying the energy-momentum tensor
we find that the total central charge of the new $T$ vanishes. This 
is due to the topological nature of the model under the analysis 
(see for example 
\ganor\ for a complete 
analyis of the BRST cohomology for $G/G$ and $G/H$ models). 

In fact, 
it is easy to see that the ghost introduced can be viewed as twisted 
fermions on the worldsheet
\eqn\aVII{
\psi_{\a} = {\xi_{\a} + \beta_{\a} \over 2}\,, ~~~~~ 
\bar \psi_{\a} = {\xi_{\a} - \beta_{\a} \over 2}\,, 
}
$$
\bar\psi^{i} ={ \eta^{i} + \bar \eta^{i} \over 2}\,,~~~~~~~
\psi^{i} ={ \eta^{i} - \bar \eta^{i} \over 2}\,,
$$
and the BRST symmetry as a twisted supersymmetry on the worldsheet. 
To compute the total central charge it is sufficient to compute the 
anomaly in the ghost current 
\eqn\aVIII{
J^{gh}(z) = - \sum_{\a} \xi_{\a} \beta^{\a} - \sum_{i} \eta^{i} \bar\eta_{i}\,.
}
Since the ghosts fields are free fields, the coefficient of the double 
pole of $J^{gh}(z) J^{gh}(w)$ is 
$N^{2} -1$, namely the dimension of the Lie group 
$SU(N)$.   For the supergroup $SU(M|N)$ one has $N^{2} + M^{2} - 
2 M \, N - 2$. 

Following the previous sections, we have to define a new BRST 
charge which leads to the correct cohomology of the theory. For 
that purpose we follow the previous section and 
we add the topological quartet formed by the commuting 
fields $(\phi^{i}, \bar\phi_{i})$ and by the anticommuting ghosts
$(\eta^{\prime i}, \bar\eta'_{i})$. They are needed to remove the 
ghosts $\eta^{i}, \bar\eta_{i}$ added in \aVI. The introduction of 
new fields might modify the central charge, but introducing 
topological quartets the total central charge remains zero. 
Nevertheless the coefficient of the double pole in 
$J^{gh}(z) J^{gh}(w)$, where 
\eqn\aVIII{
J^{gh}(z) = - \sum_{\a \in \Delta} \xi_{\a} \beta^{\a} - \sum_{i} \eta^{i} \bar\eta_{i} - 
\sum_{i} \eta^{\prime i} \bar\eta'_{i}\,.
}
changes from $N^{2} -1$ to $N^{2} + N - 2$. 

The new BRST charge is given by 
\eqn\aIX{
j_{C} = \sum_{i, \a \in \Delta_{+}} 
\bar{\eta}'_i \a^i \xi^{\a} \xi^{-\a} + \sum_{i} \bar\phi_{i} \eta^{i} \,,
}
and $j_{K}$ has to modified as follows
\eqn\aX{
j_{K} \rightarrow j_{K} + \sum_{i} \bar\phi_{i} \eta^{\prime i} \,,
}
Both currents are nilpotent and they anticommute 
$j_{K}(z) j_{C}(w) \sim 0$. 
In the present framework, we can establish a new 
conserved current 
\eqn\aX{
J^{gr} = - \sum_{i} \eta^{i} \bar\eta_{i} - 
\sum_{i} \eta^{\prime i} \bar\eta'_{i}
}
which corresponds to the assignment in \grading. Notice that 
the second BRST current $j_{C}$ contains only pieces with 
grading $-1$ and $+1$. We can clearly make any linear 
combination of the current $J^{gh}$ and $J^{gr}$. The coefficient 
of the double poles of the second charge is $2(N-1)$. Notice that the 
construction achieved so far resembles the $N=4$ embedding of the  
RNS superstrings provided in 
\bvw. In particular, for the 
N=4 formulation of the superstring, the two BRST 
charges $Q_{1}$ and $Q_{2}$ 
implement the superdiffeomorphims at the quantum 
level and restrict the Fock subspace to the small Hilbert space. 
The two currents $J_{1}$ and $J_{2}$ 
are identified with ghost and picture number. 
Both BRST charges have ghost number one, but while the second
has picture $-1$, the first one is a sum of terms with definite picture. 
This reproduces the structure outlined above. The ghost number and 
the picture number have to be identified with $J^{gh}$ and $J^{gr}$, 
and the two BRST charges with $Q_{K}$ and $Q_{C}$. We can push the 
analogy even further: the motivation to introduce a second BRST 
charge in the RNS context is the enlargement of the functional space 
from the small Hilbert space (without the zero mode of $\xi$) 
to the large Hilbert space (with $\xi_{0}$). The second BRST charge 
restricts again the space, but then one can work covariantly (namely 
with all the modes of the field $\xi$). The motivation to introduce the 
second BRST charge $Q_{C}$ in the present framework is the 
enlargement of the functional space to a space without constraints \tre. 

As outlined in the previous section, one can construct the vertex operators 
and study the spectrum. As has been already shown at the 
massless level, this model has nontrivial solutions to the equations of motion.
The conformal field theory approach should however lead to the 
analysis of the complete tower of states. A detailed study for the 
superstring will be presented elsewhere. 


\newsec{Conclusions and outlook}

In this article we have shown how to gauge the set of raising 
and lowering operators for an arbitrary Lie algebra in a covariant way. One needs to introduce 
more ghosts, and then remove their effects by a second BRST charge. 
It would be interesting to consider other cosets, for example the subset 
of all raising operators which plays a role in the derivation of harmonic superspace from 
pure spinors \GrassiXC. 
 
To complete the analysis of the conformal field theory
of the previous section, we 
should repeat the analysis given in 
 \lref\figue{
J.~M.~Figueroa-O'Farrill and S.~Stanciu,
Nucl.\ Phys.\ B {\bf 484}, 583 (1997)
[hep-th/9605111]; 
J.~M.~Figueroa-O'Farrill and S.~Stanciu,
[hep-th/9511229]; 
.~M.~Figueroa-O'Farrill and S.~Stanciu,
Nucl.\ Phys.\ B {\bf 458}, 137 (1996)
[hep-th/9506151].
} \figue\ and \lref\getzler{
E.~Getzler,
Annals Phys.\  {\bf 237}, 161 (1995)
[hep-th/9309057].
} \getzler, leading to a Kazama algebra 
\lref\Kazama{
Y.~Kazama,
Mod.\ Phys.\ Lett.\ A {\bf 6} (1991), 1321.
} \Kazama\ and, 
after adding a Koszul quartet (a 
topological gravity quartet), we should obtain an 
$N=2$ superconformal algebra \grassiWZW. 
Using the ghosts of topological gravity 
one can construct a new BRST charge to implement the reparametrization 
invariance as has been discussed in 
\lref\topgrav{
J. Labastida, M. Pernici, and E. Witten, Nucl. Phys. B {\bf 310} (1989) 258; 
D. Montano and J. Sonnenschein, Nucl. Phys. B {\bf 313} (1989) 258; 
D.~Montano and J.~Sonnenschein,
Nucl.\ Phys.\ B {\bf 324}, 348 (1989); 
R. Myers and V. Periwal, Nucl. Phys. B {\bf 333} (1990) 536; 
A. Chamseddine and D. Wyler, Phys. Lett. B {\bf 228} 75; K. Isler and 
C. Trugenberger, 
Phys.\ Rev.\ Lett.\  {\bf 63}, 834 (1989); E.~Witten, 
Nucl.\ Phys.\ B {\bf 340}, 281 (1990); 
R.~Dijkgraaf, H.~Verlinde and E.~Verlinde,
{\it Notes On Topological String Theory And 2-D Quantum Gravity,}
PUPT-1217
{\it Based on lectures given at Spring School on Strings and Quantum Gravity, Trieste, Italy, Apr 24 - May 2, 1990 and at Cargese Workshop on
Random Surfaces, Quantum Gravity and Strings, Cargese, France, May 28 - Jun 1, 1990}
} \topgrav, but in addition we expect to need another BRST charge to recover the 
observables of the gravity sector. For this purpose we intend to 
use the formalism developed in this paper. 

\vskip .5cm
\hskip -.7cm  {\bf Acknowledgments}
\vskip .5cm

We would like to thank W. Siegel for useful comments on non-minimal 
quartets in string theory and R. Stora for  discussions of topological 
quantum field theory. This work was partly funded by NSF Grant PHY-0098527. 
PAG thanks L. Castellani and A. Lerda for discussions and financial support. 
P.A.G. thanks the Theory Division  at CERN where this work has been 
completed. 

\vskip 1cm 

\appendix{A}{$SU(3)$ as an example}

We apply our results to $SU(3)$ as an example. To parametrize the 
compact basis we take the usual set of matrices with the normalization 
${\rm tr}(\l_{a} \l_{b}) = 2 \delta_{a b}$, so the 
first Gell-Mann matrix is given by 
\eqn\GM { \l_{1} = \left( \matrix{ 
0 & 1 & 0 \cr 
1 & 0 & 0 \cr
0 & 0 & 0 
}\right)}
On the Cartan-Weyl basis, the raising generators 
are 
$E_{I} = {-1\over 4} (\l_{1} + i \l_{2})$, 
$E_{II} = {-1\over 4} (\l_{4} + i \l_{5})$, 
and $E_{III} = {-1\over 4} (\l_{6} - i \l_{7})$, while 
$E_{-I} = E_{I}^{\dagger} = {-1\over 4} (\l_{1} - i \l_{2})$, \dots. 
The Cartan generators are the hermitian matrices $H_{T} = {1\over 2} \l_{3}$ 
and $H_{Y} = {1\over 2} \l_{8}$. The commutation 
relations which determine $N_{\a,\b}$ read 
\eqn\CR{\eqalign{
&[E_{-III}, E_{-II}] = -{1\over 2} E_{-I}\,, ~~~~~~~~~~~~~
[E_{II}, E_{III}] = -{1\over 2} E_{I}\,, \cr
&[E_{III}, E_{-I}] = -{1\over 2} E_{-II}\,, ~~~~~~~~~~~~~~~
[E_{I}, E_{-III}] = -{1\over 2} E_{II}\,, \cr
&[E_{I}, E_{-II}] = {1\over 2} E_{III}\,, ~~~~~~~~~~~~~~~~~~~
[E_{II}, E_{-I}] = {1\over 2} E_{-III}\,, \cr
}}
One easily derives 
\eqn\CR{\eqalign{
&[E_{I}, E_{-I}] = {1\over 2} H_{T}\,, ~~~~~~~~~~~~~~~~~~~~~~~~~
[E_{II}, E_{-II}] = {1\over 4} H_{T} + {\sqrt{3}\over 4} H_{Y} \,, 
\cr
&[E_{III}, E_{-III}] = {1\over 4} H_{T} - {\sqrt{3}\over 4} H_{Y}
\,, ~~~~~~~~~
[H_{T}, E_{I}] = E_{I}\,, \cr
&[H_{Y}, E_{I}] = 0\,, ~~~~~~~~~~~~~~~~~~~~~~~~~~~~~~~~~~
[H_{T}, E_{II}] = {1\over 2} E_{II}\,, \cr
&[H_{Y}, E_{II}] = {\sqrt{3} \over 2} E_{II}\,,  ~~~~~~~~~~~~~~~~~~~~~~~~~~
\dots\,. 
}}

The normalization $N_{\a,\b} =- N_{-\a, -\b}$ is satisfied. The roots 
are $(\pm 1, 0)$ and $(\pm 1/2$, $\pm \sqrt{3}/2)$. The Cartan-Killing 
metric $g_{AB} = f_{AP}^{~~~Q} f_{BQ}^{~~~P}$ is given by 
$g_{ij} = 3 \delta_{ij}$, and $g_{\a, -\a} = 3/2$ for each root. The usual 
relation $g_{\a, -\a} \a_{i} = g_{ij} \a^{i}$ is clearly satisfied. We shall 
occasionally need $g^{ij} \b_{j} = g^{\b, -\b} \b^{i}$. (We do not 
rescale $E_{\a}$ such that $g_{\a, -\a} =1$; hence, the indices of $\a^{i}$ 
and $H^{i}$ are lowered by the matrix $g_{ij} g^{\a, -\a} = 2 g_{ij} /3$, 
but we shall never have occasion to lower indices).

The constraints in \tre\ become 
\eqn\CRR{
C^{T} ={1\over 2} \xi^{I} \xi^{-I} + {1\over 4} \xi^{II} \xi^{-II} + {1\over 4} 
\xi^{III} \xi^{-III} =0 \,, 
~~~~~~~
C^{Y} = {\sqrt{3} \over 4} \left(\xi^{II} \xi^{-II} - \xi^{III} \xi^{-III} \right)=0 \,.  
}
The BRST charge $Q_{K,0}$ in (2.2) becomes
\eqn\CRRR{
Q_{K,0} = \Big( \xi^{I} E_{I} + \dots + \xi^{-III} E_{-III}\Big) 
}
$$
+{1\over 2} 
\Big(\b_{I} \xi^{II} \xi^{III} + \b_{II} \xi^{I} \xi^{-III} - \b_{III} \xi^{I} \xi^{-II} - 
\b_{-I} \xi^{-II} \xi^{-III} - \b_{-II} \xi^{-I} \xi^{III} + \b_{-III} \xi^{-I} \xi^{II} \Big)
$$
The constraints commute with $Q_{K,0}$, as one may check by explicit computation. 

The quadratic Casimir operator is given by 
\eqn\CCA{
C_{2} = {2\over 3} \Big(
E_{I}E_{-I} + E_{II}E_{-II} + E_{III}E_{-III} + 
E_{-I}E_{I} + E_{-II}E_{II} + E_{-III}E_{III} \Big) 
}
$$
+{1\over 3} \Big(H_{T} H_{T} + H_{Y} H_{Y} \Big)\,.
$$

The square of $Q_{K,0}$ contains only constraints 
\eqn\CICCIO{
(Q_{K,0})^{2} = C^{T} H_{T} + C^{Y} H_{Y}\,.
}
Adding 
\eqn\CICCIOO{\eqalign{
& Q_{K,-1} = - \bar \eta_{T} C^{T} - \bar\eta_{Y} C^{Y} \cr
& Q_{K, 1}' = \eta^{T} H_{T} + \eta^{Y} H_{Y} 
\bar \phi_{T} \eta^{\prime T} + \bar \phi_{Y} \eta^{\prime Y}
\,, \cr
& Q_{K,1}'' = \eta^{T} \Big( {1\over 2} 
\b_{I} \xi^{I} + {1\over 2} \b_{II} \xi^{II} + 
{1\over 2} \b_{III} \xi^{III} -{1\over 2}\b_{-I} \xi^{-I} - {1\over 2} \b_{-II} \xi^{-II}
 - 
{1\over 2} \b_{-III} \xi^{-III} \Big) + \cr
& ~~~~~~ + 
{\sqrt{3} \over 2} 
\eta^{Y} \Big( \b_{II} \xi^{II} - \b_{III} \xi^{III} - 
\b_{-II} \xi^{II} + \b_{-III} \xi^{-III} \Big) 
}}
we find the nilpotent BRST operator $Q_{K}$. 

The second BRST operator $Q_{C}$ is given by 
\eqn\CRRP{
Q_{C} = \bar\eta^{\prime T} 
\Big({1\over 2} \xi^{I} \xi^{-I} + {1\over 4} \xi^{II} \xi^{-II} + {1\over 4} 
\xi^{III} \xi^{-III} \Big) +  
}
$$
~~~~~~~~~~
+{\sqrt{3} \over 4} \bar\eta^{\prime Y} 
 \left(\xi^{II} \xi^{-II} - \xi^{III} \xi^{-III} \right) 
+ \bar \phi_{T} \eta^{T} + \bar \phi_{Y} \eta^{Y}\,,
$$
and one may verify by direct computation that indeed anticommutes 
with $Q_{K}$. 

\vskip 1cm 

\appendix{B}{
The Haar measure for $SU(2)/U(1)$ from the BRST cohomology}

\vskip .5cm 

In this appendix we present an application of the formalism presented 
in the previous sections. We consider the group $SU(2)$ and 
we parametrize the matrix of the coset  $SU(2)/U(1)$ with a single complex 
vector $p_{i}$ with $i=1,2$. We assume that $p_{i}$ is normalized to 
unity and a given $u \in SU(2)/U(1)$ can be written 
as $u = (p_{i}, \e_{ij} \bar p^{j})$. 
 
Associated to the generators $E_{\pm}$ and $H$, 
we introduce the following differential operators 
\eqn\suI{
D_{+} = p_{i} \e^{ij} \p_{\bar p^{j}}\,, ~~~~~
D_{-} = \bar p^{i} \e_{ij} \p_{p_{j}}\,, ~~~~~
[D_{+}, D_{-} ] = D_{0} \equiv p_{i} \p_{i} - \bar p^{i} \p_{\bar p^{i}}
}
$$
[D_{0}, D_{+}] =D_{+}\,, ~~~~~~
[D_{0}, D_{-}] = - D_{-}\,, ~~~~~~
 $$
and the BRST charge 
\eqn\suII{
Q_{K,0} = \xi^{+} D_{+} + \xi^{-} D_{-}
}
which is nilpotent if $\xi^{+} \xi^{-} =0$. 
Acting with $Q_{K,0}$ on the vector $p_{i}$ and its conjugate 
$\bar p^{i}$ (they are treated as independent), \suII\ leads to 
\eqn\suIII{
\{Q_{K,0}, \bar p^{i} \}= \xi^{+} p_{k} \e^{ki}\,, ~~~~~~~ \{Q_{K,0}, \xi^{+}\} =0\,, 
} 
$$
\{Q_{K,0}, p_{i} \}= \xi^{-} \bar p^{k} \e_{ki}\,, ~~~~~~~ 
\{Q_{K,0}, \xi^{-}\} =0\,.  
$$

Let us introduce the homogenous forms
\eqn\suIV{
\omega^{+} = \bar p^{i} \e_{ij} d \bar p^{j}\,, ~~~~
\omega^{-} = p_{i} \e^{ij} d p_{j}\,. 
 }
They are dual of $D_{+}$ and $D_{-}$ in the sense that 
$\langle D_{\pm}, \omega^{\pm} \rangle = 1$ and 
$\langle D_{\pm}, \omega^{\mp} \rangle = 0$ when 
$\langle \p_{p_{i}}, dp_{j} \rangle = \delta^{i}_{j}$, 
$\langle \p_{\bar p_{i}}, d\bar p_{j} \rangle = \delta^{i}_{j}$, etc.
Their BRST variations are given by 
\eqn\suV{
\{Q_{K,0}, \omega^{+}\} = - 2 \xi^{+} p_{i} d\bar p^{i} + d \xi^{+}\,, ~~~~~~
\{Q_{K,0}, \omega^{-}\} =  2 \xi^{-} \bar p_{i} d p_{i} + d \xi^{-}\,, ~~~~~~
} 
where we used that $Q_{K,0}$ and the exterior derivative $d$ 
anticommute, and 
$d\xi^{+}$ and $d\xi^{-}$ have to be considered as 
the worldsheet derivatives of the ghost fields. 

With some algebra, it is easy to show that 
\eqn\suVI{
\{Q_{K,0}, \omega^{+} \omega^{-} \} = 
d \Big( \xi^{+} \omega^{-} + \omega^{+} \xi^{-} \Big)\,,
}
Then, computing the $Q_{K,0}$ 
variation of $\Big( \xi^{+} \omega^{-} + \omega^{+} \xi^{-} \Big)$, one gets
\eqn\suV{
\left\{Q_{K,0}, \Big( \xi^{+} \omega^{-} + \omega^{+} \xi^{-} \Big) \right\} = 
\xi^{+}\xi^{-} (-4 \bar p^{k} dp_{k}) - d (\xi^{+} \xi^{-}) 
}
which is zero because of the constraints $\xi^{+} \xi^{-} =0$. 
This shows that $\omega^{+} \omega^{-}$ belongs to the 
cohomology of $Q_{K,0}$ modulo $d$-exact terms, and 
satisfies the descent equations. By 
using the parametrization 
$z = {p_{1} \over p_{2}}$ and $\bar z = {\bar p^{1} \over \bar p^{2}}$ 
a simple exercise shows that 
\eqn\suVI{
\omega^{+} \omega^{-} = {dz d\bar z \over (1+|z|^{2})^{2}} \,.
}
This is the Haar measure of the coset $SU(2)/U(1)$. 


\listrefs
\bye